\documentstyle[prl,aps,psfig,twocolumn]{revtex}
\begin{document}
\twocolumn[\hsize\textwidth\columnwidth\hsize\csname
@twocolumnfalse\endcsname

\title{Chemical efficiency of reactive microflows
with heterogeneus catalysis: a lattice Boltzmann study}

\author{S. Succi$^{1,3*}$, A. Gabrielli$^{2}$, G. Smith$^{3}$, 
E. Kaxiras$^{3}$}
\address{
$^1$ Istituto di Applicazioni Calcolo,
viale Policlinico 137, 00161 - Roma, Italy}
\address{
$^2$ INFM, Dipartimento di Fisica, Universit\`a di Roma 
"La Sapienza", P.le A. Moro 2, 00185 - Roma, Italy}
\address{
$^3$ Lyman Laboratory of Physics,
Harvard University, Cambridge, USA}
\address{
$^3*$ Visiting Scholar, Lyman Lab. of Physics, Harvard University\\
}
\date{\today}
\maketitle

\begin{abstract}
We investigate the effects of geometrical micro-irregularities
on the conversion efficiency of reactive flows in narrow
channels of millimetric size. 
Three-dimensional simulations, based upon a
Lattice-Boltzmann-Lax-Wendroff code, indicate that 
periodic micro-barriers may have an appreciable 
effect on the effective reaction efficiency of the device. 
Once extrapolated to macroscopic scales, these effects can
result in a sizeable increase of the overall reaction efficiency.
\end{abstract}
]
\narrowtext

\section{Introduction}

The formulation of mathematical models and attendant simulational
tools for the description of complex phenomena involving multiple
scales in space and time represents one of the outstanding frontiers 
of modern applied physics/mathematics \cite{KAX}.
One such example of complex multiscale phenomena is the dynamics
of reactive flows, a subject of wide interdisciplinary
concern in theoretical and applied science, with several 
applications in molecular engineering, material science, 
environmental and life sciences alike.
The complexity of reactive flow dynamics 
is parametrized by three dimensionless quantities:
the {\it Reynolds number $Re=UL/\nu$}, the 
{\it Damkohler number $Da=\tau_h/\tau_c$}, 
and the {\it Peclet} number $Pe=\frac{UH}{D}$. 
Here $U$, $L$ and $H$ denote the macroscopic flow speed and 
longitudinal/transversal lengths of the flow, respectively, $\nu$
the fluid kinematic viscosity and $D$ the pollutant 
molecular diffusivity.
The quantities $\tau_c$ and $\tau_h$ represent typical timescales
of chemical and hydrodynamic phenomena. 

High Reynolds numbers are associated with turbulence, namely loss 
of coherence of the flow field in both space and time.
High Damkohler numbers imply that chemistry is much faster
than hydrodynamics, so that reactions are always
in chemical equilibrium and take place in tiny regions 
(thin flames, reaction pockets) of evolving flow configurations. 
The opposite regime (``well-stirred'' reactor) characterizes
situations where the chemistry is slow and always takes place at
local mechanical equilibrium.
Finally, high Peclet numbers imply that the transported species
stick tightly to the fluid carrier (in the limit $Pe \rightarrow \infty$
the tracer field is ``frozen-in'' within flow streamlines).
Navigation across the three dimensional $Re-Da-Pe$ parameter
space meets with an enormous variety of 
chemico-physical behaviours, ranging from turbulent combustion
to hydrodynamic dispersion and others \cite{ORAN}.
The picture gets further complicated when geometry is taken into 
account, since boundary conditions select the spatio-temporal
structures sustaining the non-linear interaction 
between the various fields.
In this work we shall deal with {\it low-Reynolds, fast-reacting flows
with heterogeneus catalysis}. In particular we wish to gain insights
into the role of geometric micro-irregularities on the 
effective rate of absorption of tracer species 
(pollutant hereafter) at catalytic boundaries.
This is a theme of broad interest, with applications in biology, 
physics, chemistry, environmental sciences and more.
It is therefore hoped that such kind of theoretical-computational 
studies may promote a better understanding of the 
complex phenomena behind these important applications \cite{CATAL}.

\section{Mathematical model of reactive microflow dynamics}

We shall deal with an {\it incompressible, isothermal flow} with 
soluted species which are transported (advect and diffuse)
by the flow and, upon reaching solid walls,
they undergo {\it catalytic chemical reactions}.
The basic equations of fluid motion are:
\begin{eqnarray}
\label{FLOW}
\partial_t \rho + div (\rho \vec{u}) = 0\\
\partial_t (\rho \vec{u}) + div ( \rho \vec{u} \vec{u})=-\nabla P 
+ div (\mu \nabla \vec{u}) 
\end{eqnarray}
where $\rho$ is the flow density, $\vec{u}$ the flow speed,
$P=\rho T$ the fluid pressure, $T$ the temperature
and $\mu=\rho \nu$ the dynamic viscosity and 
$\vec{u} \vec{u}$ denotes the dyadic tensor $u_{a}u_{b},\;a,b=x,y,z$.

\vspace{3mm}

Multispecies transport with chemical reactions is described
by a set of generalized continuity-diffusion equations:

\begin{equation}
\label{TRA}
\partial_t C_s + div (C_s \vec{u}_s) =  
div [D_s \nabla (C_s/\rho)] + \dot \Omega_s
\end{equation}
where $C_s$ denotes the mass density of the generic $s$-th species,
$D_s$ its mass diffusivity and
$\dot \Omega_s$ is a surface-chemical reaction term to
be detailed shortly.
In the following we indicate with the subscripts $w$ and $g$ the 
``wall'' (solid) and ``gas'' in contact with the wall respectively.
According to Fick's law, the outgoing (bulk-to-wall) {\em diffusive 
mass flux} is given by:
\begin{equation} 
\vec{J}_{g\rightarrow w}=-D \nabla C^g|_w\,.
\end{equation} 
Upon contact with solid walls, the transported species react according
to the following empirical rate equation 
(the species index being removed for simplicity): 
\begin{equation}
\label{CWALL}
\dot \Omega \equiv \frac{d C^{w}}{dt} = \Gamma_w -K_c C^w
\end{equation}
where the wall-flux is taken in the simple linear form:
\begin{equation}
\label{WALL}
\Gamma_{w}=K_w (C^g-C^w)
\end{equation}
where $K_w$ is the wall to/from fluid mass transfer rate and 
$K_c$ is the chemical reaction rate dictating
species consumption once a molecule is absorbed by the wall.
The subscripts $w$ and $g$ mean ``wall'' (solid) and ``gas'' in a contact
with the wall respectively.
The above rate equation serves as a dynamic boundary condition
for the species transport equations, so that
each boundary cell can be regarded as a microscopic
chemical reactor sustained by the mass inflow from the fluid.
In the absence of surface chemical reactions
the species concentration in the solid wall would pile up in time,
up to the point where no outflow would occurr, 
a condition met when $C^g=C^w$.
Chemistry sets a time scale for this pile-up and fixes the
steady-state mass exchange rate.
At steady state we obtain:
\begin{equation}
C^w = \frac{K_w}{K_w+K_c} C^g
\end{equation}
hence
\begin{equation}
\Gamma^w = \frac{C^g}{\tau_w+\tau_c} 
\end{equation}
where $\tau_w=1/K_w$ and $\tau_c=1/K_c$.
These expressions show that finite-rate chemistry ($K_c>0$) 
ensures a non-zero steady wall outflux of pollutant.
At steady state, this mass flow to the catalytic wall
comes into balance with chemical reactions, thus fixing
a relation between the value of the wall-gradient concentration 
and its normal-to-wall gradient:

$\left\|D \partial_\perp C^g|_w\right\|= p \; C^g/(\tau_c+\tau_w)$, 

where $\partial_\perp$ means the normal to the perimeter component 
of the gradient and $p$ is the perimeter (volume/area) of the reactive cell.
This is a mixed Neumann-Dirichlet boundary condition and
identifies the free-slip length of the tracer as
$l_s=D(\tau_w+\tau_c)/p$.

\section{The computational method}

The flow field is solved by a lattice Boltzmann method 
\cite{MZ,HSB,LBE,LBGK} while the multispecies transport 
and chemical reactions are handled with a variant of the 
Lax-Wendroff method \cite{JCP}.
A few details are given in the following.

\subsection{Lattice Boltzmann equation}

The simplest, and most popular form of lattice Boltzmann equation
(Lattice BGK, for Bahtnagar, Gross, Krook)
\cite{LBGK}, reads as follows:

\begin{equation} \label{LBGK} 
f_i(\vec{x}+\vec{c}_i,t+1)-f_i(\vec{x},t)= -\omega[f_i-f_i^e](\vec{x},t)
\end{equation} 

where $f_i(\vec{x},t) \equiv f(\vec{x},\vec{v}=\vec{c}_i,t)$ is a discrete population
moving along the discrete speed $\vec{c}_i$. 
The set of discrete speeds must be chosen in such a way as to guarantee
mass, momentum and energy conservation, as well as rotational invariance.
Only a limited subclass of lattices qualifies.
In the sequel, we shall refer to the nineteen-speed lattice consisting
of zero-speed, speed one $c=1$ (nearest neighbor connection), and
speed $c=\sqrt 2$, (next-nearest-neighbor connection).
This makes a total of $19$ discrete speeds, $6$ neighbors, $12$
nearest-neighbors and $1$ rest particle ($c=0$).
The right hand side of (\ref{LBGK}) represents 
the relaxation to a local equilibrium $f_i^e$
in a time lapse of the order of $\omega^{-1}$. 
This local equilibrium is usually taken in the form
of a quadratic expansion of a Maxwellian:
\begin{equation}
f_i^e = \rho \left[1+\frac{\vec{u} \cdot \vec{c}_i}{c_s^2}
+\frac{\vec{u}\vec{u} \cdot (\vec{c}_i \vec{c}_i-c_s^2 I)}{2 c_s^4}
\right]
\end{equation}
where $c_s$ is the sound speed and $I$ denotes the identity.

Once the discrete populations are known, fluid density
and speed are obtained by (weighted) sums over
the set of discrete speeds:
\begin{equation}
\rho = m\sum_i f_i,\;\;\;
\rho \vec{u}=m\sum_i f_i \vec{c}_i
\end{equation}
LBE was historically derived as the one-body kinetic equation
resulting from many-body Lattice Gas Automata, but it can mathematically
obtained by standard projection upon Hermite polynomials
of the continuum BGK equation and subsequent evaluation of the
kinetic moment by Gaussian quadrature \cite{HE}. 
It so happens that the discrete
speeds $\vec{c}_i$ are nothing but the Gaussian knots, showing
that Gaussian integration achieves a sort of automatic
``importance sampling'' of velocity space which allows to
capture the complexities of hydrodynamic flows by means of 
only a handful of discrete speeds.
The LBE proves a very competitive tool for the numerical studies
of hydrodynamic flows, ranging from complex flows in porous
media to fully developed turbulence. 

\subsection{Modified Lax-Wendroff scheme for species transport}

Since species transport equation is linear in the species
concentration, we can solve it on a simple 
6-neighbors cubic lattice. 
Within this approach, each species is associated with
a species density $C_s$, which splits into six separate
contributions along the lattice links.

With these preparations, the transport operator 
in $3$ dimensions reads as follows (in units of $\Delta t = 1)$):
\begin{equation}
C_s (\vec{x},t)=\sum_{j=0}^{6} p_j(\vec{x}-\vec{c}_j,t-1) C_{s}
(\vec{x}-\vec{c}_j,t-1)
\end{equation}
The index $j$ runs over $\vec{x}$ and its {\it nearest-neighbors} 
(hence simpler than the LBE stencil) spanned by the vectors
$\vec{x}+\vec{c}_j$, $j=1,6$, $j=0$ being associated
with the node $\vec{x}$ itself.
The break-up coefficient $p_j$ represents the probability that
a particle at $\vec{x}_j \equiv \vec{x}-\vec{c}_j$ at time $t-1$
moves along link $j$ to contribute to $C_s(\vec{x})$ at time $t$.
For instance in a one dimensional lattice
the exact expression of these coefficients
(in lattice units $\vec{c}_j = \pm 1,\;\;j=1,2$, $\Delta t = 1$)
is:
\begin{eqnarray}
p_i (x \pm 1,t-1)& =& \frac{1 \mp u'}{2} + D'_s, \;\;\;i=1,2 \\
p_0(x,t-1) &=& -2 D'_s
\end{eqnarray}

where $u'=(u+\rho^{-1}\;\partial_x \rho)$ is the effective
speed, inclusive of the density gradient component, and
$D'_s = D_s (1-u'^2)/2$ is the effective diffusion, the
square $u'$ dependence being dictated by arguments of
numerical stability.

\subsection{Multiscale considerations}

The simulation of a reactive flow system is to
all effects a {\it multi-physics} problem
involving four distinct physical processes:
{\em
\begin{enumerate}
\item Fluid Motion (F)         
\item Species Transport (T)
\item Fluid-Wall interaction (W)  
\item Wall Chemical Reactions (C)
\end{enumerate}
}
Each of these processes is characterized by its own
timescale which may differ 
considerably from process to process depending on the 
local thermodynamic conditions. 
Loosely speaking, we think of $F$ and $T$ as to macroscopic
phenomena, and $W$ and $C$ as of microscopic ones.
The relevant fluid scales are the advective and 
momentum-diffusive time, and
the mass-diffusion time of the species respectively:
\begin{equation}
\begin{array}{ll}
\tau_A = L/U,\\
\tau_{\nu}=H^2/\nu,
\end{array}
\end{equation}
where $L,H$ are the length and height of the fluid domain.
The relevant time scales for species dynamics are: 
\begin{equation}
\begin{array}{lll}
\tau_D=H^2/D,\\
\tau_w=K_w^{-1},\\
\tau_c=K_c^{-1}
\end{array}
\end{equation}
As discussed in the introduction, they
define the major dimensionless parameters
\begin{eqnarray}
Re=UH/\nu \equiv \tau_A/\tau_{\nu},\\
Pe=UH/D \equiv \tau_A/\tau_{D},\\
Da_c=\tau_c/\tau_A,\; Da_w=\tau_w/\tau_A
\end{eqnarray}

To acknowledge the multiscale nature in time of the problem,
a subcycled time-stepper is adopted. 
This is organized as follows.
The code ticks with the hopping time of the fluid populations
from a lattice site to its neighbors $dt=dx/c=1$. 
Under all circumstances $dt$ is much smaller than both diffusive
and advective fluid scales in order to provide a faithful
description of fluid flow. 
Whenever $dt$ exceeds the chemical time-scales (high Damkohler regime),
{\it fractional time-stepping}, i.e. subcycling of the 
microscopic mechanisms, namely chemical-wall transfer is performed.
This means that the chemical and wall transfer operators are performed
$dt/\tau_c$,$dt/\tau_w$ times respectively at each fluid cycle.
As it will be appreciated shortly, since the flow solver ticks at
the sound speed, the present microflow simulations proceed in very
short time steps, of the order of tens of nanoseconds.
This means that they can be in principle coupled to mesoscopic
methods, such as kinetic Monte Carlo, affording a more
realistic description of the fluid-wall interactions.
In particular, a Kinetic Monte Carlo update of a single
boundary cell could proceed in parallel with a corresponding
hydrodynamic treatment of the entire pile of fluid cells on
top of the wall. 
The flip side of the medal is that in order to draw quantitative 
conclusions at the scale of the macroscopic devices
a two-three decade extrapolation is required. 
This commands a robust scaling argument.

\section{Catalytic efficiency: qualitative analysis}

Ideally, we would like to synthetize a universal 
functional dependence of the
catalytic efficiency as a function
of the relevant dimensionless numbers and
geometrical design parameters:
\begin{equation}
\eta=f(Re,Da,Pe;\bar g).
\end{equation}
where $\bar g$ represents a vector of geometric parameters
characterizing the boundary shape.
The question is to assess the sensitivity
of $\eta$ to $\bar g$  and possibly find an
optimal solution (maximum $\eta$) within the given parameter space.
Mathematically, this is a complex non-linear functional optimization
problem for the geometrical parameters.
We find it convenient to start from a simple-and yet representative-baseline
geometry as an ``unperturbed'' zero order approximation, which
is easily accessible either analytically or numerically.
Perturbations to this baseline situation can then be parametrized as
``topological excitations'' on top of the geometrical ``ground state''.
In the present study, the unperturbed geometry is a straight channel of size
$L$ along the flow direction and $H \times H$ across it.
Perturbations are then defined as micro-corrugations in the bottom wall
of the form $z=h(x,y)$, $h \equiv 0$ being the smooth-wall unperturbed case.
In this work, the perturbation is taken in the form of 
delta-like protrusions (barriers) 
$h(x,y,z)=\sum_i h_i \delta(x-x_i)$. 

From a macroscopic point of view the device efficiency is defined
as amount of pollutant consumpted per unit mass injected: 
\begin{equation}
\label{ETA}
\eta = \frac{\Phi_{in}-\Phi_{out}}{\Phi_{in}} 
\end{equation}
where 
\begin{equation}
\label{PHI}
\Phi(x)=\int [uC] (x,y,z) dy dz
\end{equation}
is the longitudinal mass flow of the pollutant at section $x$.
The in-out longitudinal flow deficit is of course equal to the
amount of pollutant absorbed at the catalytic wall, namely
the normal-to-wall mass flow rate: 
\begin{equation}
\label{TGAMMA}
\Gamma= \int_S \vec{\gamma}(x,y,z) \cdot d \vec{S}
\end{equation}
where the flux consists of
both advective and diffusive components:
\begin{equation}
\label{GAMMA}
\vec{\gamma}= \vec{u} C - D \nabla C
\end{equation}
and the integral runs over the entire absorbing surface $S$  

The goal of the optimization problem is to maximize $\Gamma$ 
at a given $\Phi_{in}$.
As it is apparent from the above expressions, this means maximizing
complex configuration-dependent quantities, such as the wall distribution
of the pollutant and its normal-to-wall gradient.
For future purposes, we find it convenient to recast the
catalytic efficiency as $\eta=1-T$, where $T$ 
is the channell {\it transmittance}
\begin{eqnarray}
T \equiv \Phi_{out}/\Phi_{in}
\end{eqnarray}

From a microscopic viewpoint, $T$ can be regarded as the probability
for a tracer molecule injected at the inlet to exit the channel
without being absorbed by the wall and consequently
it fixes the escape rate from the chemical trap.
Roughly speaking, in the limit of
fast-chemistry, this is controlled by the ratio 
of advection to diffusion timescales. 
More precisely, the escape rate is high if the cross-channel
distance walked by a tracer molecule in a transit time $\tau_A$ is much
smaller than the channel cross-length $H/2$.
Mathematically:
$D \tau_A \ll H^2/4$, which is: 
\begin{equation}
Pe \gg 4 \; L/H
\end{equation}
The above inequality (in reverse) shows that in order 
to achieve high conversion
efficiencies, the longitudinal aspect ratio 
$L/H$ of the device has to scale linearly with the Peclet number. 

\subsection{The role of micro-irregularities}

We now discuss the main qualitative 
effect of geometrical roughness on the above picture from
a microscopic point of view, i.e. trying to resolve flow
features at the same scale of the micro-irregularity.

In the first place, geometric irregularities provide a potential
enhancement of reactivity via the sheer 
increase of the surface/volume ratio.
Of course, how much of this potential is actually realized depends
on the resulting flow configuration.

Here, the fluid plays a two-faced role.
First, geometrical restrictions lead to local fluid acceleration,
hence less time for the pollutant molecules to migrate from the
bulk to the wall before being convected away by the mainstream
flow. This effect, usually negligible for macroscopic flows, may become
appreciable for micro-flows with $h/H \simeq 0.1$ (like in actual
catalytic converters), $h$ being the typical geometrical
micro-scale of the wall corrugations.
Moreover, obstacles shield away
part of the active surface (wake of the obstacle) where the fluid
circulates at much reduced rates (stagnation) so that
less pollutant is fed into the active surface.
The size of the shielded region is proportional to
the Reynolds number of the flow.
On the other hand, if by some mechanism the flow proves capable
of feeding the shielded region, then efficient absorption is restored
simply because the pollutant is confined by recirculating patterns and has
almost infinite time to react without being convected away.
The ordinary mechanism to feed the wall is molecular 
diffusion/dispersion, which is usually rather slow
as compared to advection.
More efficient is the case where the flow 
develops local micro-turbulence which may 
increase bulk-to-wall diffusive transport
via enhanced density gradients and attendant 
density jumps $C^g-C^w$: 
\begin{equation} 
\Gamma^{tur}_{w}=-[w'C]_{w}
\end{equation} 
where $w'$ is the normal-to-wall microturbulent velocity fluctuation. 
This latter can even dominate the picture whenever turbulent fluctuations
are sufficiently energetic, a condition met when the 
micro-Peclet number exceeds unity:
\begin{equation}
Pe_h=\frac{w'h}{D} \gg 1
\end{equation}
where $h$ is the typical geometrical micro-scale.
Given this complex competition of efficiency-promoting and efficiency-degrading
interweaved effects it is clear that assessing which type
of micro-irregularities can promote better efficiency is a non-trivial
task.

\subsection{Efficiency: analytic and scaling considerations}

For a smooth channel, the steady state solution of the
longitudinal concentration field away from the inlet
boundary factors into the product
of three independent one-dimensional 
functions: $C(x,y,z)=X(x)Y(y)Z(z)$. 
Replacing this ansatz into the steady-state version of
the equation (\ref{TRA}) we obtain: 
\begin{equation}
\begin{array}{lll}
\label{XYZ}
X(x)=X_0 e^{-x/l}\\
Y(y)=Y_0\\
Z(z)=Z_0 \cos (z/l_{\perp})
\end{array}
\end{equation}
with the longitudinal and cross-flow absorption lengths
related via: 
\begin{equation}
\label{LLP}
l= l_{\perp}^2 \frac{\bar U}{D}                    
\end{equation}
where $\bar U$ is the average flow speed 
\begin{equation}
\bar U(x)=\sum_{y,z} u(x,y,z) C(x,y,z)/\sum_{y,z} C(x,y,z)
\end{equation}
Note that the profile along the spanwise coordinate $y$ remains
almost flat because we stipulate that
only the top and bottom walls host catalytic reactions.

To determine the cross-flow absorption length $l_{\perp}$
we impose that along all fluid cells in a contact with the wall,
the diffusive flux is exactly equal to fluid-to-wall outflow,
namely:
\begin{equation}
\label{LPERP}
\frac{C}{l_{\perp}^2}= \frac{C_g}{\tau} \frac{2}{N_z}                    
\end{equation}
where $\tau$ the effective absorption/reaction time scale,
\begin{equation}
\frac{1}{\tau} \simeq \frac{1}{\tau_D}+\frac{1}{(\tau_c+\tau_w)}\,,
\end{equation}
and $N_z=H^2$ is the number of cells ($dx=1$ in the code)
in a cross-section $x=const.$ of the channel.
Therefore the factor $2/N_z$ is the fraction of reactive cells
along any given cross-section $x=const.$ of the channel.

The form factor $C_g/C$ is readily obtained by the third
of Eq.~(\ref{XYZ}) which yields 
\begin{equation}
\frac{C_g}{C} \simeq \cos(H/2 \l_{\perp})
\end{equation}
Combining this equation with Eq.~(\ref{LPERP}) we obtain
a non-linear algebraic equation for $l_{\perp}$:
\begin{equation}
\label{LAMBDA}
\lambda^{-2} \; \cos (\lambda/2) = \frac{D \tau}{H^2} \frac{N_z}{2}
\end{equation}
where we have set $\lambda \equiv H/l_{\perp}$.
For each set of parameters this equation can be easily solved numerically
to deliver $l_{\perp}$, hence $l$ via the Eq.~(\ref{LLP}).

Given the exponential dependence along the streamwise
coordinate $x$, the efficiency can then be estimated as:
\begin{equation}
\label{ETA0}
\eta_0 \simeq 1-e^{-L/l} 
\end{equation}

Note that in the low absorption limit
$L\ll l$, the above relation reduces to $\eta_0 \simeq L/l$, meaning 
that doubling, say, the absorption length 
implies same efficiency with a twice shorter catalyzer.
In the opposite high-absorption limit, $L\gg l$, 
the relative pay-off becomes increasingly less significant.

\subsection{Corrugated channel: Analytical estimates}

Having discussed tha baseline geometry, we now turn 
to the case of a ``perturbed'' geometry.
Let us begin by considering a single barrier of height $h$.
The reference situation is a smooth channel at high Damkohler
with $\eta_0=1-e^{-L/l}$.
We seek perturbative corrections in the smallness parameter 
$g \equiv h/H$, the coupling-strength to geometrical perturbations.
The unperturbed wall-flux is
\begin{equation}
\Gamma_0 \simeq 2D \frac{C_h}{h} LH\,
\end{equation}
where $C_h$ is the concentration at the tip of the barrier calculated
in the smooth channel.
Therefore $C_h/h$ is an estimate of the normal-to-wall 
diffusive gradient.               
The geometrical gain due to extra-active wall surface is
\begin{equation}
\Gamma_1 \simeq C_h u_h hH
\end{equation}
where
\begin{equation}
u_h \simeq 4U_0(g-g^2)
\label{uh}
\end{equation}
is the average longitudinal flow speed in front of the barrier along
a section $x=const.$.
The shadowed region of size $w$ in the wake of 
the obstacle yields a contribution
\begin{equation}
\Gamma_2 \simeq a \; D \frac{C_h}{h} wH
\end{equation}
where $a$ is a measure of the absorption activity 
in the shielded region. 

Three distinctive cases can be identified:

\begin{itemize}
\item $a=0$: The wake region is totally deactivated, absorption zero.
\item $a=1$: The wake absorption is exactly the same as for unperturbed flow
\item $a>1$: The wake absorption is higher than with unperturbed flow 
(back-flowing micro-vortices can hit the rear side of the barrier)
\end{itemize}

Combining these expressions we obtain the following compact expression:

\begin{equation}
\label{DETA}
\frac{\delta \eta}{\eta_0}=
\frac{\Gamma_1+\Gamma_2-\Gamma_2(h=0)}{\Gamma_0} \simeq
\frac{A}{2} \frac{h}{H} Re_h [Sc+K\;(a-1)]
\end{equation}
where $A=H/L$ is the aspect ratio of the channel
and $Sc=\nu/D$ is the Schmidt number 
(fluid viscosity/tracer mass diffusivity)
and the wake length can be estimated as 
$w/h = K Re_h$ with $K \simeq 0.1$. 

The above expression shows a perturbative (quadratic) correction
in $h$ over the unperturbed (smooth channel situation). 
However, since the effective absorption in the shielded region is affected by
higher order complex phenomena, the factor $a$ may itself
exhibit a non-perturbative dependence on $h$, so that departures 
from this quadratic scaling should not come as a surprise.
Apart from its actual accuracy, we believe expressions like (\ref{DETA})
may provide a qualitative guideline to estimate the efficiency
of generic/random obstacle distributions $[x_i,h_i]$:
In particular, they should offer a semi-quantitative insights into
non-perturbative effects due to non-linear fluid interactions
triggered by geometrical micro-irregularities.

\section{Application: reactive flow over a microbarrier}

\begin{figure}
\centerline{\psfig{file=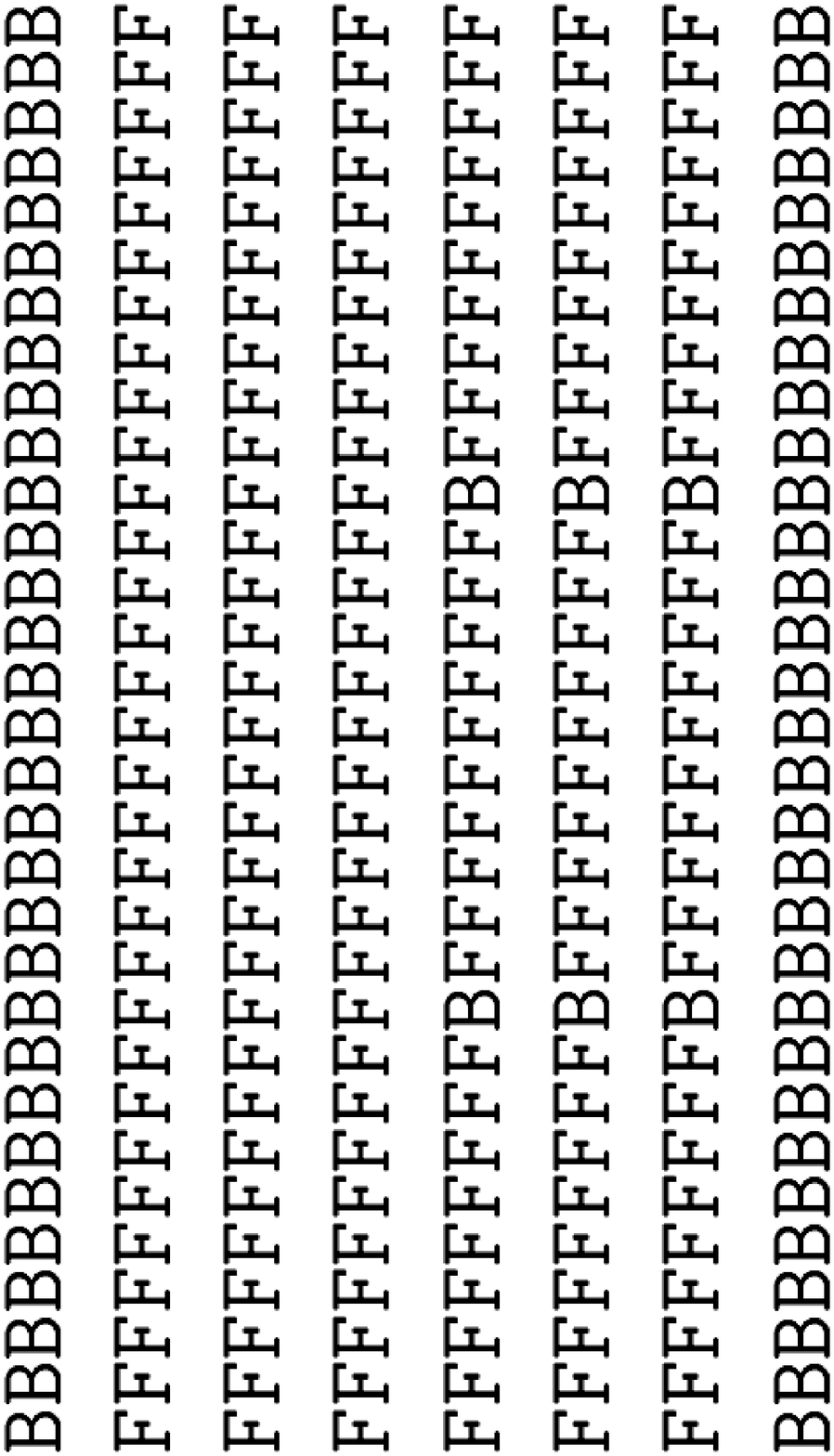,height=6cm,angle=-90}}
\caption{Sketch of the of a section at $y=const.$ of a
typical channel with two microbarriers. 
Two barriers of height $h=3$ a distance $s=10$ apart: F=fluid, B=buffer.}
\label{fig1}
\end{figure}

The previous computational scheme has been applied
to a fluid flowing in a millimeter-sized box of
of size $2 \times 1 \times 1$ millimeters along the $x,y,z$ directions
with a pair of perpendicular barriers of 
height $h$ a distance $s$ apart on the bottom wall (see Fig. \ref{fig1}
for a rapid sketch).

The single-barrier set up corresponds to the limit $s=0$.
The fluid flow carries a passive pollutant, say an 
exhaust gas flow, which is absorbed at the channel walls
where it disappears due to heterogeneus catalysis. 
The flow is forced with a constant volumetric force
which mimics the effects of a pressure gradient.
The exhaust gas is continuously injected 
at the inlet, $x=0$, with a flat profile across the channel
and, upon diffusing across the flow, it
reaches solid walls where it gets trapped and subsequently
reacts according to  a first order catalytic reaction:
\begin{equation}
\label{CO2}
C+A \rightarrow P
\end{equation}
where $A$ denote an active catalyzer and $P$ the reaction products.

The initial conditions are:
\begin{eqnarray}
&&C(x,y,z)=1,\;\;\;x=1\\
&&C(x,y,z)=0,\;\;\;\mbox{elsewher}e\\
&&\rho(x,y,z)=1\\
&&u(x,y,z) = U_0,\;\; v(x,y,z)=w(x,y,z)=0
\end{eqnarray}

The pollutant is continuously injected at the inlet and 
released at the open outlet, while 
flow periodicity is imposed at the inlet/outlet boundaries.
On the upper and lower walls, the flow speed is forced
to vanish, whereas the fluid-wall mass exchange is
modelled via a mass transfer rate equation of the form
previously discussed.

We explore the effects of a sub-millimeter
pair of barriers of height $h$ a distance $s$ apart
on the bottom wall. 
The idea is to assess the effects of the interbarrier 
$height$, $h$, and interbarrier separation $s$ on the chemical efficiency.
Upon using a $80 \times 40 \times 40$ computational grid, we
obtain a lattice with $dx=dy=dz=0.0025$ ($25$ microns), and 
$dt=c_s\; dx/V_s \simeq 50 \; 10^{-9}$ ($50$ nanoseconds).
Here we have assumed a sound speed $V_s=300$ m/s and 
used the fact that the sound speed is 
$c_s=1/\sqrt 3$ in lattice units.
Our simulations refer to the following values (in lattice units):
$U_0 \simeq 0.1-0.2$, $D=0.1$, $\nu=0.01$, $K_c=K_w=0.1$. 
This corresponds to a diffusion-limited scenario:
\begin{equation}
\tau_c = \tau_w = 10 < \tau_A \simeq 800 < \tau_D= 16000 < \tau_{\nu}=160000
\end{equation}
or, in terms of dimensionless numbers:
\begin{equation}
Pe \simeq 40,\;\;\;Re \simeq 400,\;\;\;Da>80 
\end{equation}

As per the interbarrier separation, we consider the following
values: $h/H=0.2$ and $s/L=0,1/8,1/4,1/2$, and
$h/H=0.05,0.1,0.2$ for $s/L=0$.
For the sake of comparison, the case of a smooth wall 
($s=0,h=0$) is also included.

The typical simulation time-span is $t=32000$ time-steps, namely
about $1.6$ milliseconds in physical time, corresponding
to two mass diffusion times across the channel.
The physico-chemical parameters given above are not intended
to match any specific experimental condition, 
but rather to develop a generic intuition 
for the interplay of the various processes 
in action under the fast chemistry assumption.

\subsection{Single barrier: effects of barrier heigth}

We consider a single barrier of height $h$ placed
in the middle of the bottom wall at $x=L/2,z=0$.
With the above parameters we may estimate the reference
efficiency for the case of smooth channel flow.
With $\bar U \simeq 0.1$, and $\tau=20$, we obtain
$l \simeq 200$, hence $\eta_0 \simeq 0.5$. 

\begin{figure}
\centerline{\psfig{file=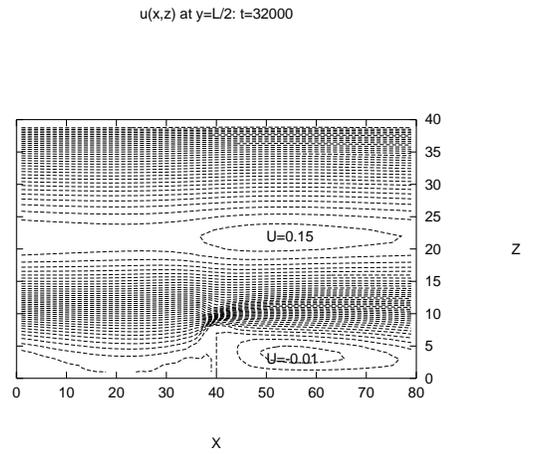,height=7cm,angle=-90}}
\caption{Typical two-dimensional cut of the flow pattern with a single barrier
of heigth $h=8$. Streamwise flow speed in the plane $y=H/2$.}
\label{fig2}
\end{figure}

\begin{figure}
\centerline{\psfig{file=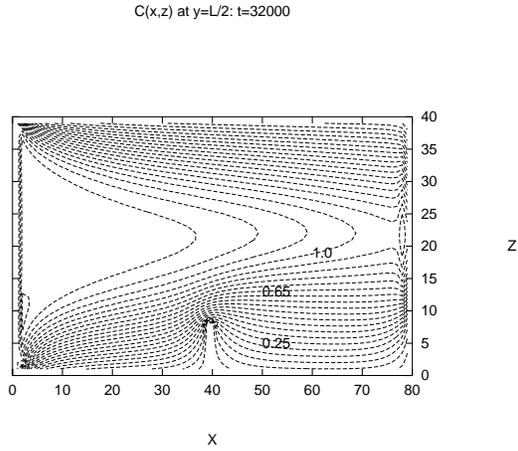,height=7cm,angle=-90}}
\caption{Concentration isocontours with a single barrier of heigth $h=8$.}
\label{fig3}
\end{figure}

\begin{figure}
\centerline{\psfig{file=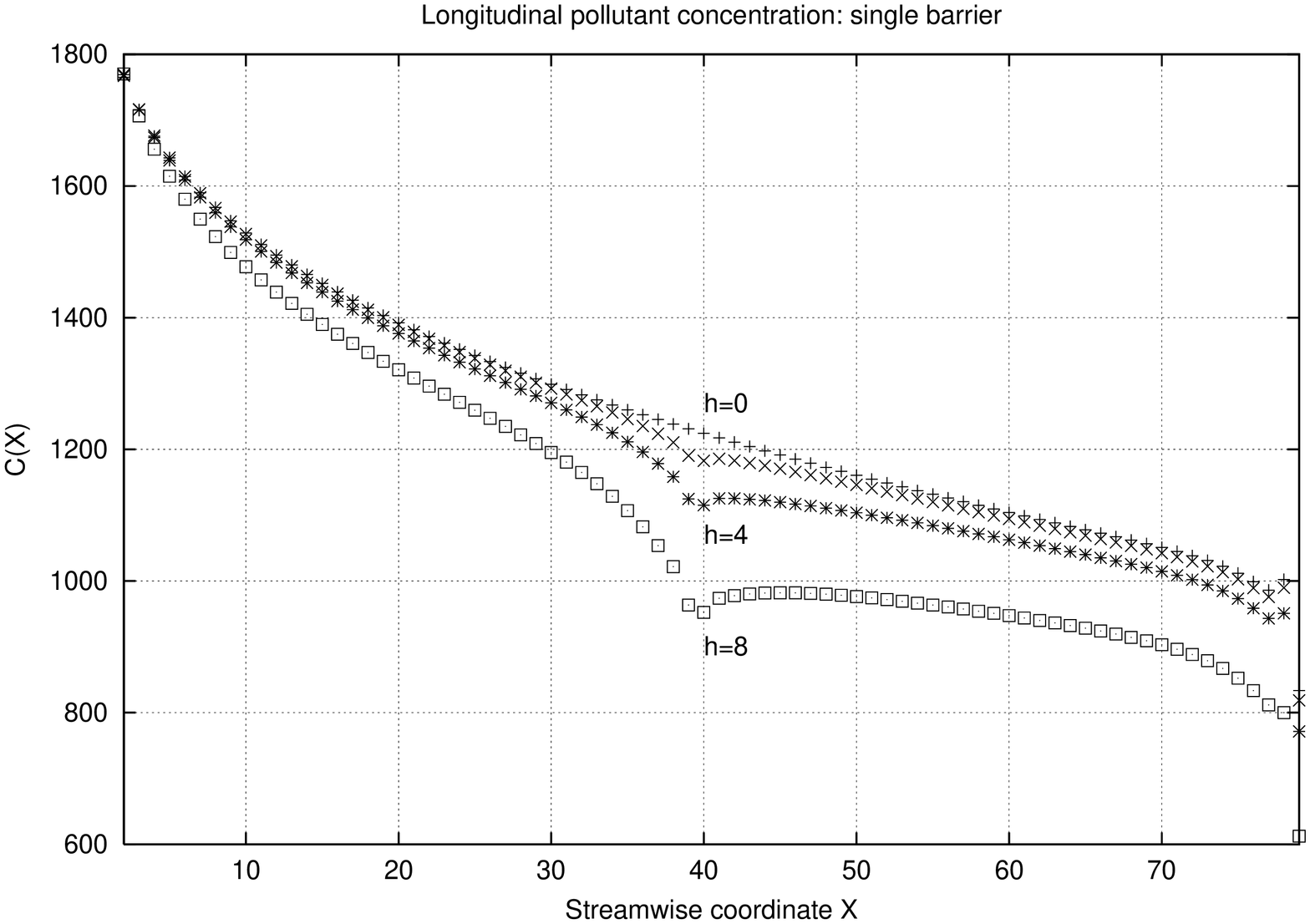,height=6.5cm,angle=-90}}
\caption{Integrated longitudinal concentration $C(x)$ of the pollutant 
with a single barrier of height $h=8$ after $32000$ steps.}  
\label{fig4}
\end{figure}

A typical two-dimensional cut of the flow pattern
and pollutant spatial distribution in the section
$y=H/2$ is shown in Figs.~\ref{fig2} and \ref{fig3}, which refer to the 
case $h=8,s=0$ ($h/H=0.1,s/L=0.0$). 
An extended (if feeble) recirculation pattern is
well visible past the barrier.
Also, enhanced concentration gradients in correspondence of
the tip of the barrier is easily recognized from Fig.~\ref{fig3}.
A more quantitative information is conveyed by Fig.~\ref{fig4}, where the
integrated longitudinal concentration of the pollutant:
\begin{equation}
C(x)=\sum_{y,z} C(x,y,z)
\end{equation}
is presented for the cases $h=0,2,4,8$ (always with $s=0$).
The main highlight is a substantial reduction of the pollutant
concentration with increasing barrier height.
This is qualitatively very plausible since 
the bulk flow is richer in pollutant and consequently the
tip of the barrier ``eats up'' more pollutant than the lower region. 
In order to gain a semi-quantitative estimate of the
chemical efficiency, we measure the
the pollutant longitudinal mass flow:
\begin{equation}
\Phi(x) = \sum_{y,z} [Cu](x,y,z) 
\end{equation}

The values at $x=1$ and $x=L$ define the efficiency according 
to Eq.~(\ref{ETA}) (to minimize finite-size effects
actual measurements are taken at $x=2$ and $x=70$). 

\begin{figure}
\centerline{\psfig{file=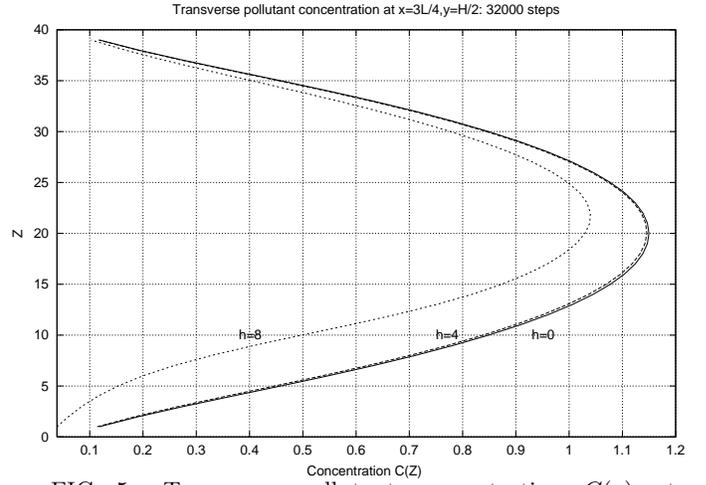,height=6.5cm,angle=-90}}
\caption{Transverse pollutant concentration $C(z)$ at $x=3L/4$ and $y=H/2$.
Single barrier of varying height. The four curves for each
of the three different heigths are taken at 
$t=3200,6400,29800,32000$.}
\label{fig5}
\end{figure}

The corresponding results are shown in Table \ref{tab1}, 
where subscript $A$ refers
to the analytical expression (\ref{DETA}) with $a=1$.
These results are in a reasonable agreement
with the analytical estimate Eq.~(\ref{DETA}) taken at 
$a=1$ (same absorption as the smooth channel).
However, for $h=8$ the assumption $a=1$ overestimates
the actual efficiency, indicating that the shielded region
absorbs significantly less pollutant than in the smooth-channel scenario.
Indeed, inspection of the transversal concentration profiles
(Fig.~\ref{fig5}) along the chord $x=3L/4,y=H/2$ reveals
a neat depletion of the pollutant in the wake region.
This is the shielding effect of the barrier.

\begin{figure}
\centerline{\psfig{file=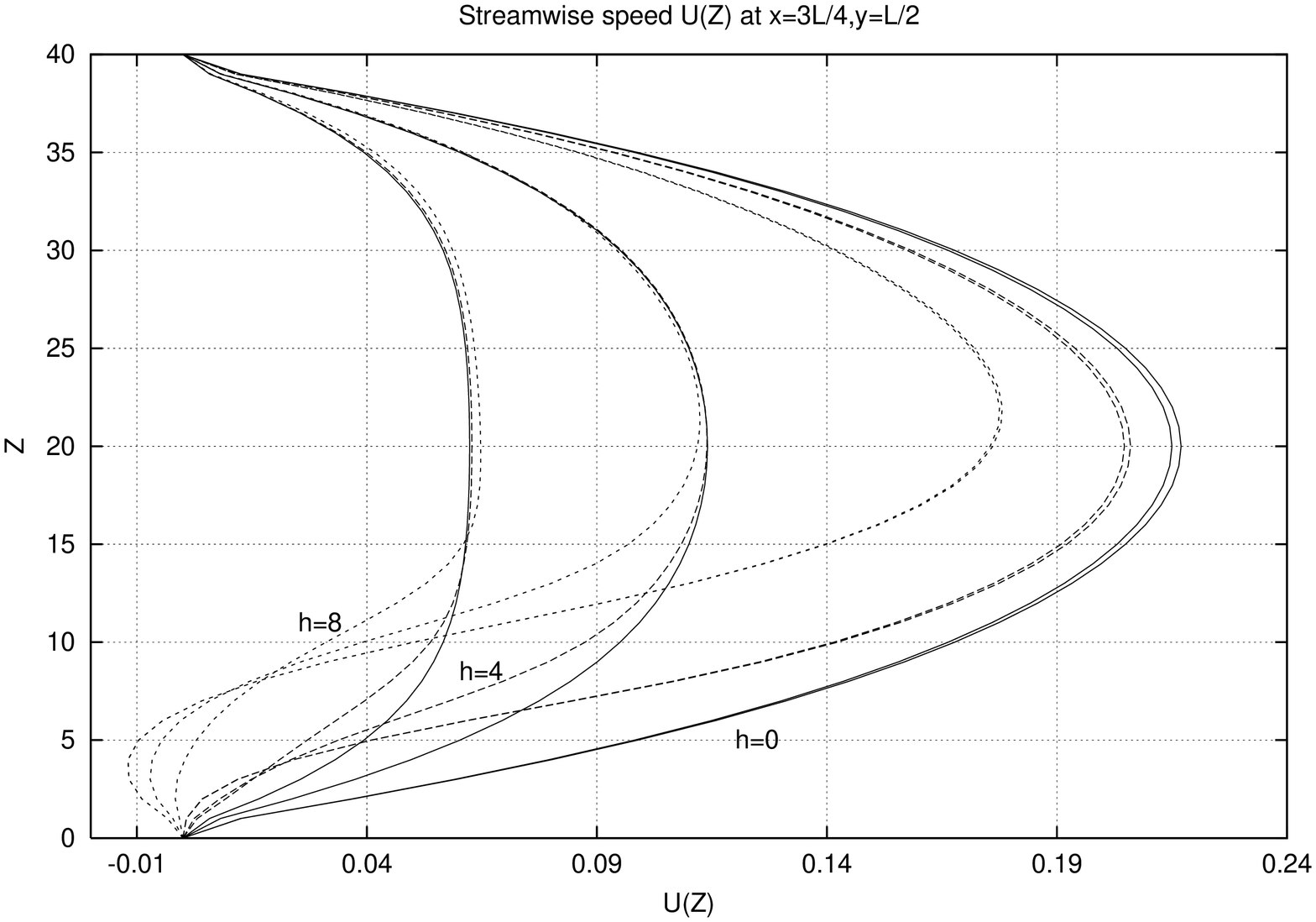,height=6.5cm,angle=-90}}
\caption{Time evolution of the transversal streamwise
speed $u(z)$ at $x=3L/4$ and $y=L/2$. 
Single barrier of varying height.}
\label{fig6}
\end{figure}

\begin{figure}
\centerline{\psfig{file=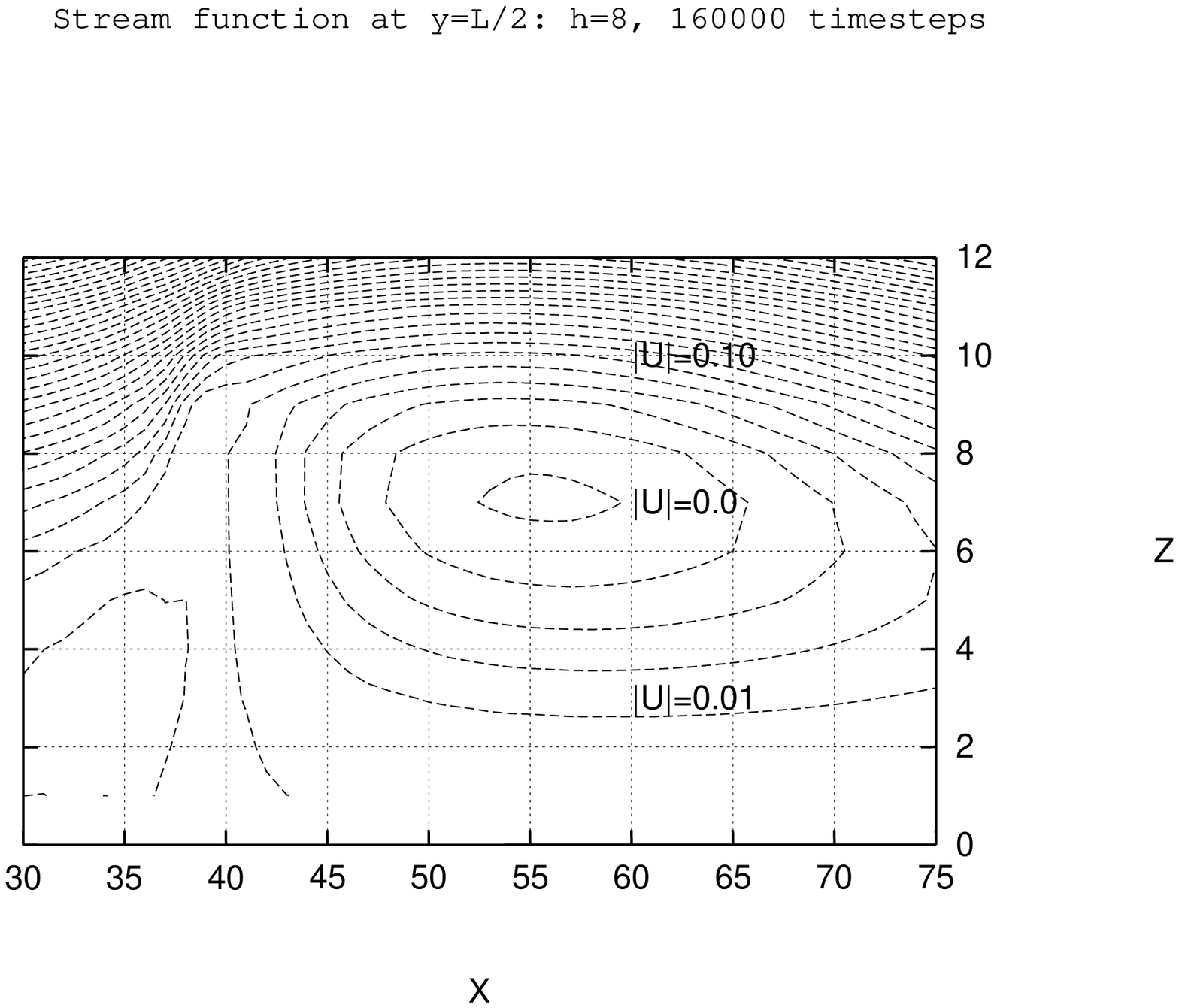,height=7cm,angle=-90}}
\caption{Blow-up of the streamlines of the flow field past a 
barrier of height $h=8$ located at $x=40$. The velocity direction
in the closed streamlines of the vortex is clockwise.}
\label{fig7}
\end{figure}

Besides this efficiency-degrading effect, the barrier also promotes 
a potentially beneficial flow recirculation, which is well visible
in Figs.~\ref{fig6} and \ref{fig7}. 
Figure \ref{fig6} shows the time evolution of the
streamwise velocity $u(z)$ in the mid-line $x=3L/4,\,y=H/2$.
It clearly reveals that recirculating backflow
only sets in for $h=8$, and also shows that the 
velocity profile gets very close to steady state.
A blow-up of the recirculating pattern in the near-wall
back-barrier region is shown in Fig.~\ref{fig7}.
However these recirculation effects are feeble (the intensity
of the recirculating flow is less than ten percent of the
bulk flow) and depletion remains the dominant mechanism.
In fact for $h=8$ the measured local Peclet number is of the order 
$0.01 \cdot 8/0.1=0.8$, seemingly too small to promote appreciable
micro-turbulent effects.
In passing, it should be noticed that raising the barrier
height has an appreciable impact on the bulk flow as well,
which displays some twenty percent reduction
due to mechanical losses on the barrier.

Finally, we observe that the measured efficiency is 
smaller than the theoretical $\eta_c$ for smoth channel. 
This is due to the fact that the flow $\Phi(x=2)$ is
significantly enhanced by the imposed inlet flat 
profile $C(z)=1$ at $x=1$ (as well visible in Fig.~\ref{fig4}).
Leaving aside the initial portion of the channel, 
our numerical data are pretty well fitted by an exponential 
with absorption length $l=160$, in a reasonable agreement
with the theoretical estimate $l \simeq 200$ obtained
by solving Eqs.~(\ref{LLP}) and (\ref{LPERP}).

\subsection{Effects of barrier separation}

\begin{figure}
\centerline{\psfig{file=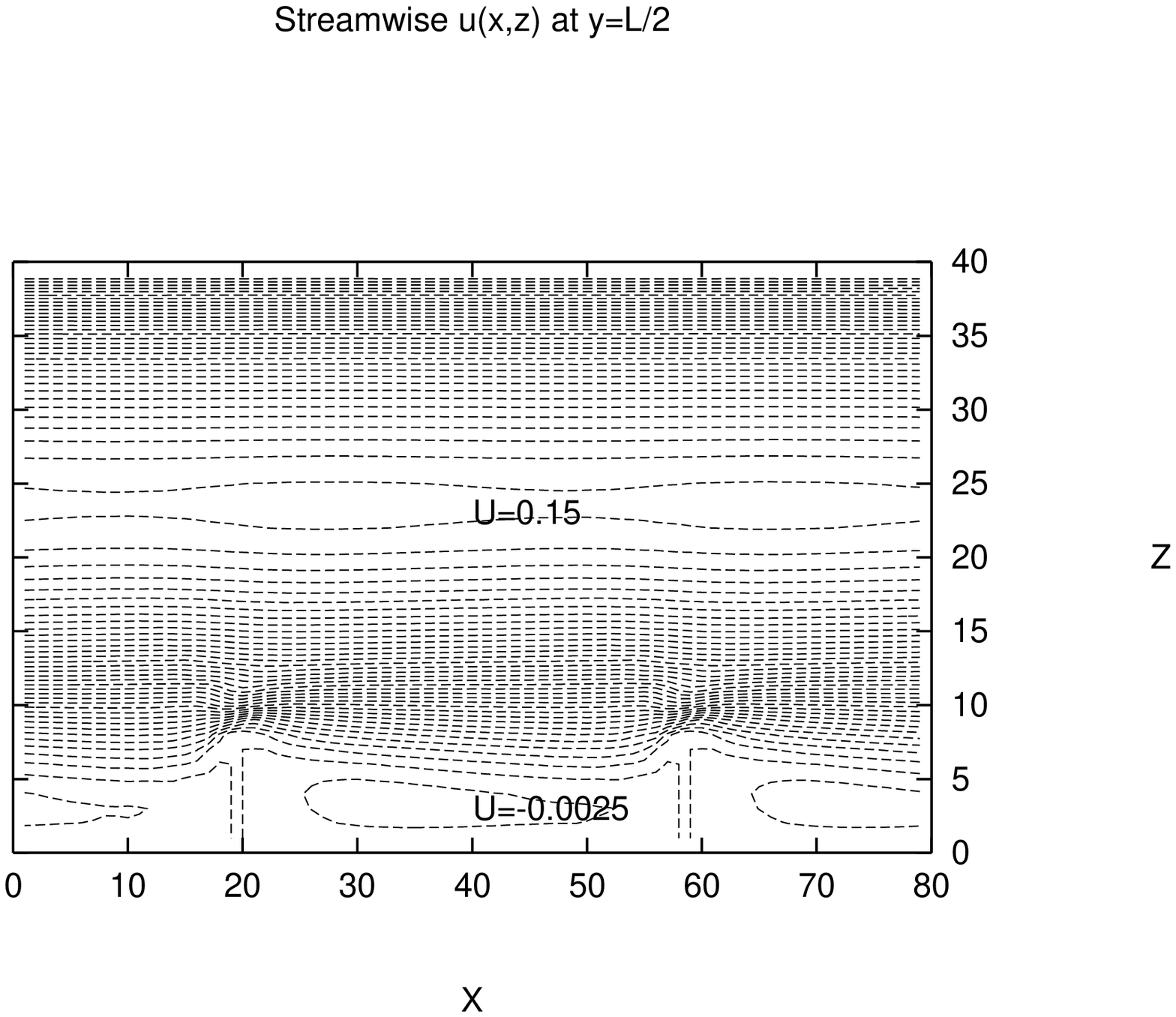,height=6.5cm,angle=-90}}
\caption{Isocontours of the streamwise flow speed with two barriers
with $h=8,s=20$ at $t=32000$.}
\label{fig8}
\end{figure}

\begin{figure}
\centerline{\psfig{file=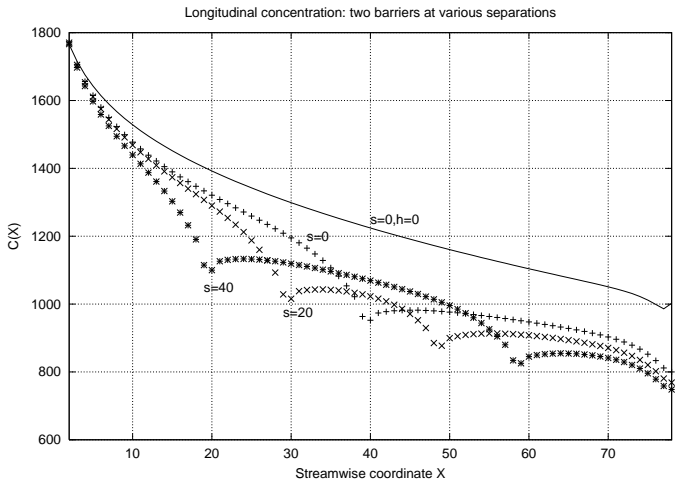,height=6.5cm,angle=-90}}
\caption{Longitudinal concentration $C(x)$ for $h=8$, $s=0,20,40$ all
at $t=32000$.}
\label{fig9}
\end{figure}

Next we examine the effect of interbarrier separation.
To this purpose, three separations $s=10,20,40$ 
symmetric around $x_0=L/2$ are been considered. 
A typical two-barrier flow pattern with $s=40$
is shown in Fig.~\ref{fig8}.
From this picture we see that even with
the largest separation $s=40$, the second barrier is still
marginally in the wake of the first one.
As a result, we expect it to suffer seriously from the
aforementioned depletion effected produced by the first 
barrier. This expectation is indeed confirmed by the
results reported in Table \ref{tab2}.
These results show that, at least on the microscopic scale, the 
presence of a second barrier does not seem
to make any significant difference, regardless of its separation
from the first one. As anticipated, the
most intuitive explanation is again shadowing: the first
barrier gets much more ``food'' than the second one, which is left
with much less pollutant due to the depletion effect 
induced by the first one.
Inspection of the longitudinal pollutant concentration
(Fig.~\ref{fig9}) clearly shows that the first barrier, regardless
of its location, ``eats up'' most of the pollutant (deficit
with respect to the upper-lying smooth-channel curve is almost
unchanged on top of the second barrier).
Of course, this destructive interference is expected to go away
for ``well-separated'' barriers with $s\gg w$.
Indeed, the ultimate goal of such investigations
should be to devise geometrical set-ups 
leading to {\it constructive interference}.
This would require much larger and longer simulations which are
beyond the scope of the present work.

\subsection{Effects of barrier height on a longer timescale}

Since the previous simulations only cover a fraction of the
global momentum diffusion time, one may wonder how
would the picture change by going to longer time scales
of the order of $H^2/\nu$.
Longer single-barrier simulations, with $t=160,000$, up to $10$ 
diffusion times, namely about $15$ milliseconds, provide the results
exposed in Table \ref{tab3}.

\begin{figure}
\centerline{\psfig{file=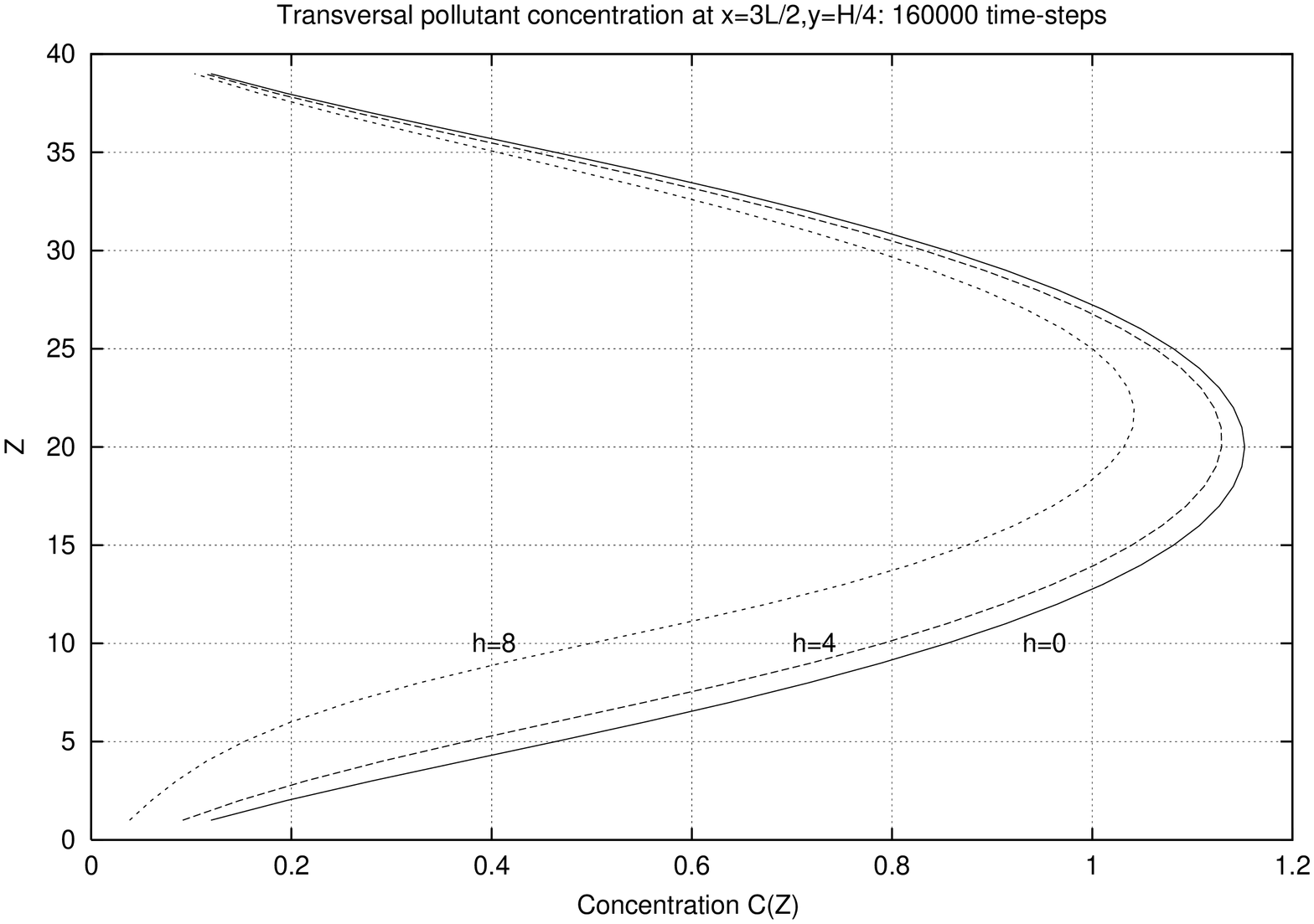,height=6.5cm,angle=-90}}
\caption{Integrated longitudinal concentration $C(x)$ of the pollutant 
with a single barrier of height $h=8$ after $160000$ steps.}  
\label{fig10}
\end{figure}

\begin{figure}
\centerline{\psfig{file=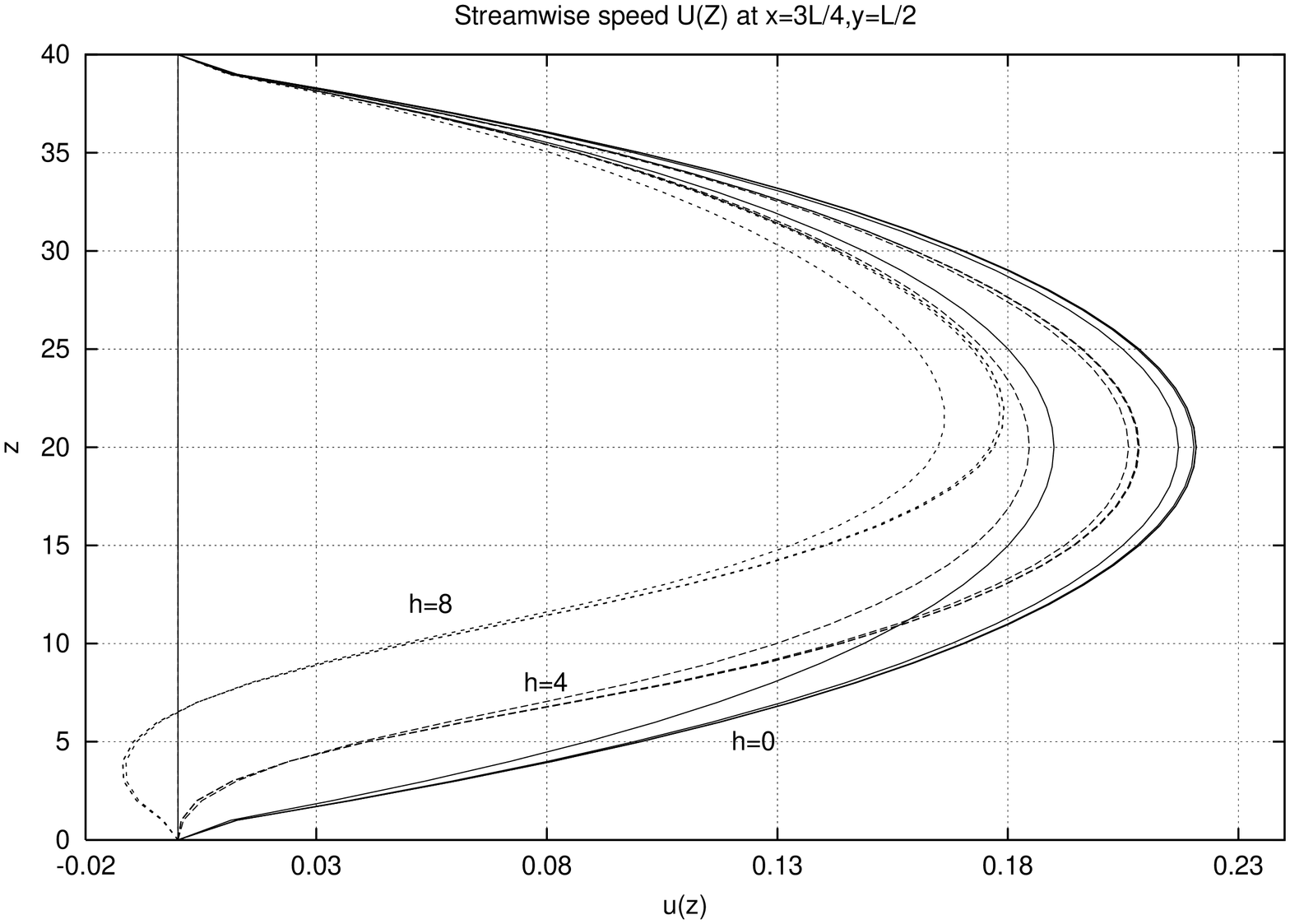,height=6.5cm,angle=-90}}
\caption{Time evolution of the transversal streamwise
speed $u(z)$ at $x=3L/4$ and $y=L/2$ after $160000$ steps. 
Single barrier of varying height.}
\label{fig11}
\end{figure}

We observe that the quantitative change is very minor, just a small
efficiency reduction due to a slightly higher flow speed.
Indeed, the spatial distribution of the pollutant does not
show any significant changes as compared to the shorter simulations. 
and a similar conclusion applies to the flow pattern (see Figs.~\ref{fig10} 
and \ref{fig11}).
This is because in a Poiseuille flow, the fluid gets quickly to, say,
$90$ percent of its total bulk speed (and even quicker to its near-wall
steady configuration), while it takes much longer to attain
the remaining ten percent. 
Since it is the near-wall flow configuration which matters mostly
in terms of a semi-quantitative estimate of the chemical efficiency,
we may conclude that the simulation span can be contained to within a fraction
of the global momentum equilibration time.

\section{Upscaling to macroscopic devices}

\begin{figure}
\centerline{\psfig{file=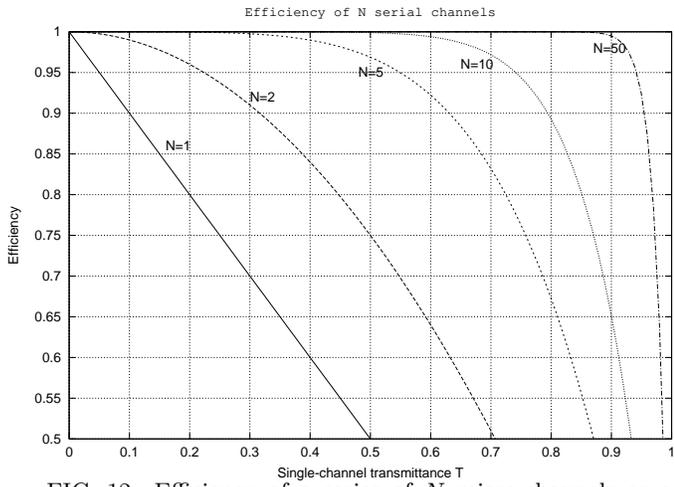,height=6.5cm,angle=-90}}
\caption{Efficiency of a series of $N$ micro-channels as a function
of the single-channel transmittance.}
\label{fig12}
\end{figure}

It is important to realize that even tiny improvements on the
microscopic scale can result in pretty sizeable cumulative
effects on the macroscopic scale of 
the real devices, say $10$ centimeters.
Assuming for a while the efficiency of an array
of $N$ serial micro-channels can be estimated simply as
\begin{equation}
\label{BOLD}
\eta_N = 1 -T^N\,,
\end{equation}
it is readily recognized that even low single-channel
efficiencies can result in significant efficiencies
of macroscopic devices with $N=10-100$ (see Fig.~\ref{fig12}). 
In particular, single-channel transmittances 
as high as $90$ percent can lead to appreciable 
macroscopic efficiencies, around $60$ percent, when just 
ten such micro-channels are linked-up together. 
Such a sensitive dependence implies that extrapolation
to the macroscopic scales, even when successfull 
in matching experimental data \cite{CORRO,SAE},
must be taken cautiously.
In fact, the above expression (\ref{BOLD}) represents of course
a rather bold upscaling assumption.
As a partial supporting argument, we note that
unless the geometry itself is made self-affine (fractal walls
\cite{SAP}), or the flow develops its own intrinsic scaling
structure (fully developed turbulence), 
the basic phenomena should remain controlled by a 
single scale $l$, independent of the device size $L$.
Since both instances can be excluded for the present work, extrapolation
to macroscopic scales is indeed conceivable.
Nonetheless, it is clear a tight sinergy between computer simulation
and adequate analytical scaling theories is in great demand to
make sensible predictions at the macroscopic scale.

\section{Conclusions}

This work presents a very preliminary exploratory study 
of the complex hydro-chemical phenomena 
which control the effective reactivity of catalytic 
devices of millimetric size.
Although the simulations generally confirm qualitative expectations
on the overall dependence on the major physical parameters, they
also highlight the existence of non-perturbative effects, such as
the onset of micro-vorticity in the wake of geometrical obstrusions,
which are hardly amenable to analytical treatment. 
It is hoped that the flexibility of the present computer tool, as combined
with semi-analytical theories, can be of significant help in developing 
semi-quantitative intuition about the subtle and fascinating interplay between 
geometry, chemistry, diffusion and hydrodynamics in the design of 
chemical traps, catalytic converters and other related devices.

\section{Acknowledgements}

Work performed under NATO Grant PST.CLG.976357.
SS acknowledges a scholarship
from the Physics Department at Harvard University.

\newpage
\begin{table}
\begin{tabular}{|c|c|c|c|} \hline
Run  & $h/H$ &   $\eta$  & $\frac{\delta \eta}{\eta},\frac{\delta \eta_A}
{\eta_A}$\\ \hline
R00  &0      &    0.295  & 0.00      \\ \hline
R02  &1/20   &    0.301  & 0.02,0.025\\ \hline
R04  &1/10   &    0.312  & 0.06,0.10 \\ \hline
R08  &2/10   &    0.360  & 0.22,0.40 \\ \hline
\end{tabular}
\caption{Single barrier at $x=40$: the effect of barrier height.}
\label{tab1} 
\end{table}

\begin{table}
\begin{tabular}{|c|c|c|} \hline
Run & $s/L$ &   $\eta$  \\ \hline
R00 &0      &  0.30  \\ \hline
R08 &1/8    &  0.36  \\ \hline
R28 &2/8    &  0.37  \\ \hline
R48 &4/8    &  0.375 \\ \hline
\end{tabular}
\caption{Two barriers of height $h=8$: Effect of interseparation $s$.}
\label{tab2}
\end{table}

\begin{table}
\begin{tabular}{|c|c|c|c|} \hline
Run  & $h/H$ &   $\eta$  & $\frac{\delta \eta}{\eta},\frac{\delta \eta_A}{\eta_A}$\\ \hline
L00  &0     &   0.290 & 0,0        \\ \hline
L02  &0/20  &   0.296 & 0.02,0.025 \\ \hline
L04  &1/10  &   0.307 & 0.06,0.10  \\ \hline
L08  &2/10  &   0.360 & 0.24,0.40  \\ \hline
\end{tabular}
\caption{$s=0$, $h=0,4,8$: 10 mass diffusion times}
\label{tab3}
\end{table}

\end{document}